 \documentclass[final,5p,times,twocolumn]{elsarticle}

\usepackage{amssymb}
\usepackage{amsmath}
\usepackage{gensymb}
\usepackage{pdfpages}
\usepackage{dblfloatfix}
\usepackage[hidelinks]{hyperref}
\usepackage{comment}
\usepackage{tikz}

\graphicspath{{./figures/}}

\journal{Elsevier}
\begin{document}

\begin{frontmatter}



\title{High-resolution multi-reflection time-of-flight mass spectrometer\\ for exotic nuclei at IGISOL}


\author[inst1]{V.A.~Virtanen}
\author[inst1]{T.~Eronen}
\author[inst1]{A.~Kankainen}
\author[inst1]{O.~Beliuskina}
\author[inst1]{Z.~Ge}
\author[inst1,inst2]{R.P.~de~Groote}
\author[inst1]{A.~Jokinen}
\author[inst1]{M.~Mougeot}
\author[inst1,inst3]{A.~de~Roubin}
\author[inst1]{J.~Ruotsalainen}
\author[inst1]{J.~Sar\'en}
\author[inst1]{A.~Takkinen}

\affiliation[inst1]{University of Jyvaskyla Department of Physics, Accelerator laboratory, Jyvaskyla, P.O. Box 35, (YFL) FI-40014, Finland}
\affiliation[inst2]{Instituut voor Kern- en Stralingsfysica, KU Leuven, B-3001, Leuven, Belgium}
\affiliation[inst3]{organization = {Université de Caen Normandie, ENSICAEN, CNRS/IN2P3, LPC Caen UMR6534},
                    addressline  = {F-14000 Caen},
                    country = {France}}


\begin{abstract}
A Multi-Reflection Time-of-Flight Mass Spectrometer (MR-ToF-MS) has been commissioned at the Ion-Guide Isotope Separator On-Line (IGISOL) facility. It consists of six electrode pairs that form a nearly energy-isochronous potential and a pulsed drift-tube to trap the ions between the electrodes. Time-of-flight peak widths down to 22~ns full-width at half-maximum and mass-resolving powers of $\approx 1.5\times 10^5$ within 20~ms have been demonstrated. The obtained time-focus and mass-resolving power depend sensitively on the trapping energy, energy spread and the number of revolutions of the ions. The mass-resolving power is affected by the temporal and energy spread of the ions entering the MR-ToF-MS, and fluctuations in the electrode voltages due to temperature variations. The longitudinal emittance corresponding to the temporal and energy spread of $^{39}$K is estimated to be 175~eVns based on the data, close to the expected 186(10)~eVns. The time-of-flight temperature sensitivity is found to be -5.55(30) ppm/K. In addition to atomic mass measurements of short-lived exotic nuclides, the MR-ToF-MS can be used as a fast mass separator for various other experiments at IGISOL and as an ion counter for laser spectroscopy and yield measurements.
\end{abstract}




\end{frontmatter}


\section{Introduction}
\label{sec:intro}
Today, on-line radioactive ion beam facilities face rapidly increasing contamination ratios while mapping the properties of increasingly exotic nuclei across the nuclear chart. Nuclei produced by the same reaction mechanisms as the nucleus of interest can severely hinder measurements by creating a vast background of ions from these unwanted species. Furthermore, as experiments often aim to study nuclei farther away from stability, the diminishing half-lives of nuclei constrain the time available to manipulate the ion beam. A method that can separate ion species quickly and with good resolution is therefore needed.

The atomic mass provides a way to determine the nuclear binding energy, which reflects the interactions between nucleons. This information is useful e.g. for modelling the astrophysical nucleosynthesis of elements, for probing nuclear structure and fundamental symmetries (see e.g. Refs.~\cite{Lunney2003,Blaum2006,Eronen2016,Dilling2018,Clark2023}). Moreover, the mass provides a way to separate ion species and thus purify samples of exotic ions. 

Currently, Penning traps are the most precise mass spectrometers; however, trouble dealing with isobarically contaminated samples and the relatively long time it takes to measure the masses limit their reach \cite{Wolf2013}. The typically used method to select masses in a Penning trap, the mass-selective buffer gas cooling method \cite{Savard1991}, provides a mass-resolving power 
\begin{equation}
    R = \frac{m}{\Delta m}
\end{equation}
of around $10^5$ in roughly 100~ms \cite{Eronen2016} leaving some exotic nuclides out of the reach of Penning traps. Isobaric contamination causes problems not only for mass spectrometry experiments but also for many other, such as laser- and decay spectroscopy experiments.

In the last decade, Multi-Reflection Time-of-Flight Mass Spectrometers (MR-ToF-MS) \cite{Wollnik1990,Casares2001,Ishida2004,Plass2013} have been shown to efficiently separate and measure the masses of exotic radioactive ion species. The MR-ToF-MS is a compact electrostatic trap that can separate and measure atomic masses cost-effectively compared to Penning traps and storage rings. The method enables mass-resolving powers of $\sim 2.0\times10^5$ to be reached in a short time of around 20~ms. Due to these advantages, almost every radioactive beam facility has, or is commissioning, an MR-ToF-MS (see e.g. \cite{Wolf2013,Schury2013,Schury2014a,Dickel2015,Chauveau2016,Liu2021,Reiter2021,Rosenbusch2023}. 

Different implementations of the MR-ToF-MS exist \cite{Plass2013}. The "Greifswald" MR-ToF-MS design, used e.g. at ISOLTRAP, utilizes a pulsed drift-electrode (in-trap lift) to inject the ions between the trap electrodes, to eject the ions and to temporally focus the ions \cite{Wolf2013}. The "Gie\ss en" MR-ToF-MS design, in use at the FRS Ion Catcher and at TITAN \cite{Dickel2015,Reiter2021}, traps the ions by pulsing the reflector electrodes. Also the MR-ToF-MS mass spectrometers at RIKEN pulse the reflector electrodes \cite{Schury2013,Schury2014a,Rosenbusch2023,Ito2018}. To further cut the amount of contaminant ions, other features, such as a mass-range selector \cite{Dickel2015}, mass-selective ion ejection \cite{Wienholtz2017}, in-MR-ToF-MS deflector \cite{Rosenbusch2023}, as well as mass-selective re-trapping \cite{Dickel2017,Reiter2021} and pulsed-mirror in-MR-ToF-MS selection \cite{Rosenbusch2023} methods exist.  


%

Here we report on the design and performance of an MR-ToF-MS commissioned at the Ion-Guide Isotope Separation On-Line (IGISOL) facility \cite{Moore2013}. The IGISOL facility produces a broad range of radioactive beams via fusion-evaporation, fission, or multinucleon transfer reactions. The MR-ToF-MS at IGISOL is a multipurpose mass spectrometer and high-resolving power mass separator for Penning-trap mass measurements, laser spectroscopy and decay spectroscopy experiments. Its design is based on the Greifswald MR-ToF-MS, also in use at ISOLTRAP \cite{Wolf2013}. We have measured the mass-resolving power and the temporal resolution $\Delta t$ of the device, including the effects of ion-optical aberrations and compare them with the model outlined in Sect.~\ref{sec:ToF}. Stability of ion ToF as a function of the trap electrode voltages and ambient temperature has also been investigated.

\section{Operating principle of an MR-ToF-MS}
\label{sec:ToF}
Mass separation by time-of-flight has been used since the 1950s \cite{Wollnik2013}. In the last few decades electrostatic ion traps have started to be used to increase the ToF by reflecting the trapped ions multiple times. The MR-ToF-MS uses this principle. Assuming the ions have been accelerated over the same potential difference, the detected ToF $t$ of an ion with mass $m$ and charge $q$ for an MR-ToF-MS can be written as
\begin{equation}
\label{eq:tof-eq}
    t = a\sqrt{\frac{m}{q}} + b,
\end{equation}
where $a$ is a constant that depends on the ion's energy along flight path and $b$ is a non-flight related constant offset. The ToF is increased by trapping the ions and reflecting them repeatedly along a closed trajectory, which can extend up to several kilometers in length. This flight takes different times for ions of different mass-to-charge ratios $m/q$. Assuming $b$ to be small, it follows from Eq. (\ref{eq:tof-eq}) by differentiation that the separation of two masses $\Delta m$ by one full-widths at half-maximum (FWHM) requires mass-resolving power $R$
\begin{equation}
    \label{eq:mrp}
    R = \frac{m}{\Delta m} = \frac{t}{2 \Delta t},
\end{equation}
where $\Delta t$ is the FWHM of the temporal peak. The typical $10^5$--$10^6$ mass-resolving powers reached by MR-ToF mass spectrometers can separate isobaric, i.e., same mass number $A$ ions or, sometimes, even isomeric states of nuclei. 

The mass-resolving power $R$ furthermore determines the relative precision of MR-ToF mass measurements. By substitution of the standard error of the mean of a Gaussian temporal distribution, Eq. (\ref{eq:mrp}) yields a relative mass precision $\frac{\sigma_m}{m}$, 
\begin{equation}
    \label{eq:simple_merr}
    \frac{\sigma_m}{m} \approx  \frac{1}{2\sqrt{2\ln{2}}}\frac{1}{\sqrt{N}R},
\end{equation}
where $N$ is the number of measured ions and $\sigma_t$ is the standard deviation of the temporal distribution. Current mass-resolving powers enable MR-ToF mass measurements with a relative mass precision of $ 10^{-7}$--$10^{-8}$.



%



The ions are ideally released into the spectrometer in temporally short, monoenergetic bursts. Radio-frequency quadrupole (RFQ) cooler-bunchers are used for this. They collect and cool the ions in a harmonic pseudopotential well, wherefrom the ions are periodically released into MR-ToF-MS. The ions are released by switching on an extraction field, which causes a small energy spread for the bunch ($\sigma_E$). Moreover, the extraction field determines the minimum temporal width $\sigma_\mathrm{th}$ of the ion bunch. As half of the ions are moving away from the buncher exit at the time of extraction, they are delayed as they turn around in the extraction field to get out of the buncher. This delay $\sigma_\mathrm{th}$ cannot be eliminated with static ion optics and limits the minimal temporal width of the ion bunch. 
The combination of these two effects is characterized by longitudinal emittance $\epsilon_\mathrm{long} = 2\pi\sigma_\mathrm{th}\sigma_E$. Together with the energy-dependence of the ion time-of-flight in the MR-ToF-MS it predetermines the achievable mass-resolving power and, in our case, can be estimated from the temperature of the buffer gas $T$, the mass-over-charge of the ions $m/q$, and the second-order coefficient of the harmonic, axial potential $V(z) = a_0z^2$: 
\begin{equation}
    \label{eq:emittance}
    \epsilon_\mathrm{long} =  2\pi k_\mathrm{B} T \sqrt{\frac{2m}{a_0 q}},
\end{equation}
where $k_b$ is the Boltzmann constant \cite{MiniBuncher}. 

%


Estimates of the mass-resolving power are typically found with series expansion of ion time-of-flight, with varying degrees of detail to account for temporal focusing \cite{Wolf2012a}, and higher than first-order, so-called aberration terms \cite{Yavor2009-dd}. Aberrations caused by the nonlinearities of the system limit the mass-resolving power, as the ions' temporal distribution is increasingly distorted by every revolution in the MR-ToF-MS. For a few mass-resolving power model examples, see e.g. Refs.~\cite{Wienholtz2017,Wolf2012a,Plas2013a}. We follow a similar path here, expanding the time-of-flight as a function of energy up to second-order to characterize the IGISOL MR-ToF-MS. Although the MR-ToF-MS's are designed to be energy-isochronous at a higher order, including isochronicity not only with respect to energy but also other parameters, such as ion angle and position at injection \cite{Yavor2009-dd}, a second-order expansion with respect to energy can nonetheless provide useful insight into the system, as will be shown in Sects. \ref{sec:ToF-U-R} and \ref{sec:tof-n}.


The time-of-flight from the buncher to the MR-ToF-MS and further to the detector can be divided to the shoot-through ToF, $t_s$, and the trapped, $n$ revolutions $t_n$ time-of-flight: 
\begin{equation}
t = t_s+t_n.    
\end{equation}
Ideally, the ion ToF would be independent of energy, i.e., energy-isochronous. In such a case $t_s = T_{s0}$ and $t_n = nT_0$, where the index shows the expansion coefficient order. However, in reality the flight time from the buncher usually depends on the ion energy, which causes the bunch to de-focus temporally. We approximate the energy-dependence of $t_s$ and $t_n$ by power series expansion. For practical purposes, we expand $t_s$ and $t_n$ up to linear and quadratic terms, respectively. We truncate the shoot-through expansion after the linear term as here we are mostly interested in the aberrations from the MR-ToF-MS. We develop the expansion about energy $E_0$ with  $\mathit{\tilde{E}} = E-E_0$, where the energies $E$ and $\tilde{E}$ are Gaussian random variables.
%
%
%
This way, the expansion of the total time-of-flight becomes
\begin{equation}
t = t_s+t_{n} =T_{s0} + T_{s1}\tilde{E}+ n(T_0 + T_1 \tilde{E} + T_2 \tilde{E}^2),
\label{eq:tof_total}
\end{equation}
where $T_{s1}$ is the first-order expansion coefficient for the shoot-through ToF, and $T_1$ and $T_2$ are the linear and quadratic coefficients of the ToF for $n$ revolutions. 
When using the pulsed drift-electrode to change the ion trapping energy in the MR-ToF-MS by a small amount, here denoted by  $\lambda$, it mostly affects $t_n$ as the trapped path in the MR-ToF-MS is far longer than the MR-ToF-MS portion of the shoot-through path. Assuming $\tilde{E}$ to be distributed around zero with standard deviation $\sigma_E$, the expectation value of time-of-flight $\langle t \rangle$ is
\begin{equation}
    \langle t \rangle = T_{s0} + n(T_0  + T_1\lambda + T_2[\sigma_E^2+\lambda^2]).
    \label{eq:mean}
\end{equation}
The overall variance $\sigma_t^2$ in time-of-flight $t$ is then:
\begin{equation}
    \sigma_t^2 =\sigma_{th}^2+\sigma_{t,\tilde{E}}^2=\sigma_{th}^2+(n[2T_2\lambda+T_1]+T_{s1})^2\sigma_E^2+ 2n^2T_2^2\sigma_E^4 \,
    \label{eq:var}
\end{equation}
where $\sigma_{t,\tilde{E}}$ is the variance of $t$ from Eq.~(\ref{eq:tof_total}). It should be noted that the term $(n[2T_2\lambda+T_1]+T_{s1})^2\sigma_E^2$ can be made to vanish by choosing the right combination of $n$ and $\lambda$ to temporally focus the ion bunch. However, the other terms of Eq. (\ref{eq:var}) cannot be eliminated in the same way.

\begin{figure*}[th!]
  \centering
  \includegraphics[width=1.0\textwidth]{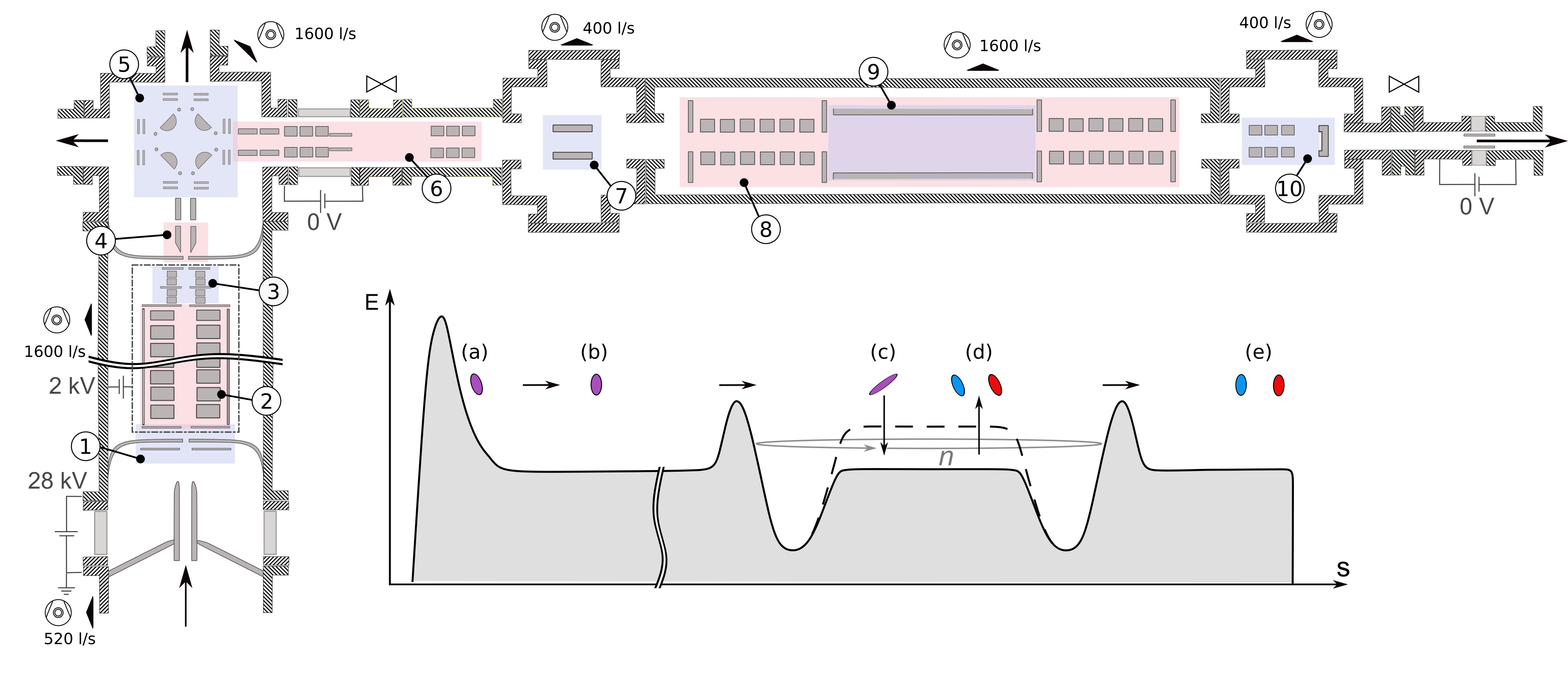}
  \caption{System schematic. Continuous ion beam enters the RFQ cooler-buncher through injection electrodes (1). The main cooling section (2) potential is closely matched with the 30~kV acceleration potential of IGISOL to efficiently capture the ions. The ions are bunched in section (3) and accelerated to 2~kV through extraction electrodes (4). The 2~kV bunches are guided to the MR-ToF-MS with a 90\degree~ electrostatic quadrupole bender (5), a double XY-steerer, a quadrupole triplet, an einzel lens (6), and an XY steerer (7). The MR-ToF-MS consist of two six-electrode stacks (8) and a central pulsed drift-tube used to inject and extract ions (9). A set of three cylindrical electrodes on the exit side of the spectrometer spatially focus the ions before the MagneToF (10). The inset displays the ions' total energy along their flight from the Mini-Buncher (a) to the detector (e). The shaded part represents the fraction of potential energy. Ions leave the Mini-Buncher (a). The ions reach the time-focus plane before the MR-ToF-MS (b) and enter the in-trap lift electrode de-focused (c). The MR-ToF-MS is tuned to focus the ions as they revolve $n$ times inside the spectrometer. When extracted (d), the separated ions reach the ion detector (e) at time-focus.}
  \label{fig:scheme}
\end{figure*}

Despite being designed to be energy-isochronous at a high order, the energy spread $\sigma_E$ of the ions nevertheless affects the time-of-flight dispersion and the mass-resolving power of the MR-ToF-MS. Using Eqs. (\ref{eq:mrp}), (\ref{eq:mean}), and (\ref{eq:var}) the mass-resolving power can be written as:
\begin{equation}
    \small
    R = \frac{T_{s0}+n(T_0  + T_1\lambda + T_2[\sigma_E^2+\lambda^2])}{ 4\sqrt{2\ln2}\cdot\sqrt{\sigma_{th}^2+(n[2T_2\lambda+T_1]+T_{s1})^2\sigma_E^2+ 2n^2T_2^2\sigma_E^4}}.
    \label{eq:mrp_fancy}
\end{equation}
Although the mass-resolving power increases as a function of the revolution number $n$ in the MR-ToF-MS, the ions also disperse in time due to the energy spread $\sigma_{E}$.  We have experimentally studied the dependence of mass-resolving power on the energy of trapped ions and number of revolutions $n$ and show in Sec. \ref{sec:performance} that the equations for $\langle t \rangle$, $\sigma_t$ and $R$ agree  well with the experimental data.



\section{Experimental setup}
At the IGISOL facility \cite{Moore2013}, the Radio-Frequency Quadrupole cooler-buncher (RFQ cooler-buncher) \cite{MiniBuncher,Nieminen2001} has been recently upgraded with a so-called Mini-Buncher to accommodate the temporal bunch width requirements of the MR-ToF-MS \cite{MiniBuncher}. With 202(11)~eVns longitudinal emittance \cite{MiniBuncher}, the upgraded buncher gives greater control over the energy- and temporal spread of the extracted ion bunches.
The system has a few distinctive features. For instance, the IGISOL RFQ cooler-buncher and MR-ToF-MS are floated on a 30-kV high-voltage platform and the ion bunches from RFQ cooler-buncher pass through a 90\degree~quadrupole bender before entering the MR-ToF-MS. In the following, we describe the operation and the main technical features of the spectrometer, schematically illustrated in Fig.~\ref{fig:scheme}.




\subsection{High-voltage platform and ion transport}

The radioactive ion beams of IGISOL are produced at 30 kV potential and accelerated to ground potential \cite{Moore2013}. The continuous ion beam is slowed down as it enters the RFQ cooler-buncher high-voltage platform slightly below the 30~kV potential of IGISOL \cite{MiniBuncher}. The ions are extracted from this potential towards the MR-ToF-MS 2~kV below the RFQ cooler-buncher potential. Since the spectrometer operates on the same high-voltage platform, fluctuations of the RFQ cooler-buncher 30~kV platform \cite{YuPGangrsky_2004} do not affect ion time-of-flight.


The ions are extracted from the RFQ cooler-buncher with the Mini-Buncher \cite{MiniBuncher}. The ions are focused and accelerated through a conical extraction lens (Fig.~\ref{fig:scheme}, (4)) to the line potential 2~kV below the RFQ cooler-buncher. An electrostatic quadrupole bender (Fig.~\ref{fig:scheme}, (5)) turns the beam 90\degree~to right towards the MR-ToF-MS beam-line. The bender consists of four main rods, each accompanied by a pair of smaller shim electrodes on their side to finely adjust the bending electric field. Additionally, horizontally focusing pairs of plates are placed on the bender entrance and exit. After the bender, the beam can be further steered both horizontally and vertically with two sets of XY-steerers and focused with an electrostatic quadrupole triplet. 
The ions then pass through another cylindrical focusing electrode (Fig.~\ref{fig:scheme}, (6)) and a final XY-steerer before reaching the MR-ToF-MS spectrometer (Fig.~\ref{fig:scheme}, (7)). 

\subsection{Vacuum system}
It is vital to run the MR-ToF-MS in ultra-high vacuum ($p < 10^{-7}$\,mbar) as any ion collision with a background particle can result in losing the ion. The MR-ToF-MS volume is pumped with a 1600 l/s turbomolecular pump and reaches $10^{-10}$~mbar-level vacuum. With the leaking helium from RFQ cooler-buncher upstream $\sim5\times 10^{-9}$~mbar pressure in the spectrometer chamber can be reached thanks to the differential pumping sections (each with 400~l/s pump) on both sides of the spectrometer.

\subsection{MR-ToF-MS and its operation}
\label{sec:MRToFoper}
The MR-ToF-MS spans approximately 81~cm and  consists of two 17~cm long stacks of mirror and lens electrodes on both sides of the spectrometer with a $47$~cm long cylindrical drift electrode between them (Fig.~\ref{fig:scheme}, (8) - (9)). The identical stacks contain six 24~mm inner diameter electrodes. The nearest electrodes to the central drift electrode, E1 and E2 spatially focus the beam. The remaining four electrodes E3-E6 reflect the ions and produce the ions' near-isochronicity with respect to kinetic energy. The identical electrode pairs are supplied with the same voltages. The electrodes are made of 316L stainless steel insulated from each other using aluminum oxide disks. The drift electrode is made of stainless steel mesh to improve pumping of the enclosed region.

The inset of Fig.~\ref{fig:scheme} shows a schematic of the MR-ToF-MS' axial potential. The potential was initially optimized with SIMION \cite{Dahl2000} and then by measuring the mass-resolving power with bunched ions while iterating the voltages applied to the electrodes by ISEG EHS 8240X module. See Table~\ref{tab:mirrorV} for their typical values, which were tuned for ions that have been trapped with 1~keV kinetic energy by pulsing the central drift-tube. 

\begin{table}[th]
    \centering
    \begin{tabular}{ll}
    \hline\hline
    Electrode & Voltage (V)\\\hline\hline
    E1 & -1871.790 \\
    E2 & -1655.055\\
    E3 & 110.890\\
    E4 & 905.900\\
    E5 & 1014.113\\
    E6 & 1469.512\\ \hline
    \end{tabular}
    \caption{Typical voltages applied to the mirror electrode pairs. E1 are next to drift tube whereas E6 are the farthest out.}
\label{tab:mirrorV}
\end{table}

The ions are trapped in the MR-ToF-MS by slowing them down by a 1~kV potential applied to the in-trap lift electrode with a Spellman MPS series +5 kV module. Once the ions are inside the electrode, the electrode is quickly switched with a Behlke HTS-61-03-GSM switch to beam-line potential leaving the ions trapped in the MR-ToF-MS. The ions then fly in the spectrometer until they have separated, as shown in the inset of Fig.~\ref{fig:scheme}. Finally, the 1~kV potential is reapplied to the in-trap lift with the switch, and the re-gained potential energy allows the ions to accelerate out of the spectrometer to the MagneToF detector \cite{MagneToF} (Fig.~\ref{fig:scheme}, (10)). 

\begin{figure}[th!]
  \centering
 \includegraphics[width=0.45\textwidth]{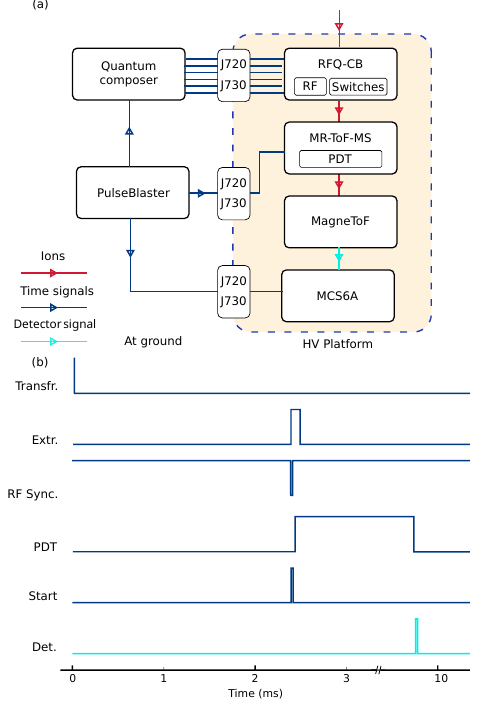}
  \caption{(a) Timing hardware and (b) typical timing sequence for the MR-ToF-MS at IGISOL. The PulseBlaster signals a Quantum Composer that controls the transfer and extraction of the ions from the RFQ cooler-buncher. Synchronized RF keeps the same phase at the extraction of each cycle. PulseBlaster sends a start signal to the MCSA6A TDC and controls the in-trap lift voltage switching in the pulsed-drift tube (PDT). The ions hit a MagneToF detector as they fly out of the MR-ToF-MS and cause a detection signal (Det.) to propagate to the TDC. The ToF is found from the difference of the start and detection signals. The signals from the PulseBlaster and Quantum Composer enter the high-voltage (HV) platform via J720 and J730 optical-electrical converters from Highland Technologies.}
  \label{fig:timings}
\end{figure}
\begin{figure}[th!]
  \centering
  \includegraphics[width=0.5\textwidth]{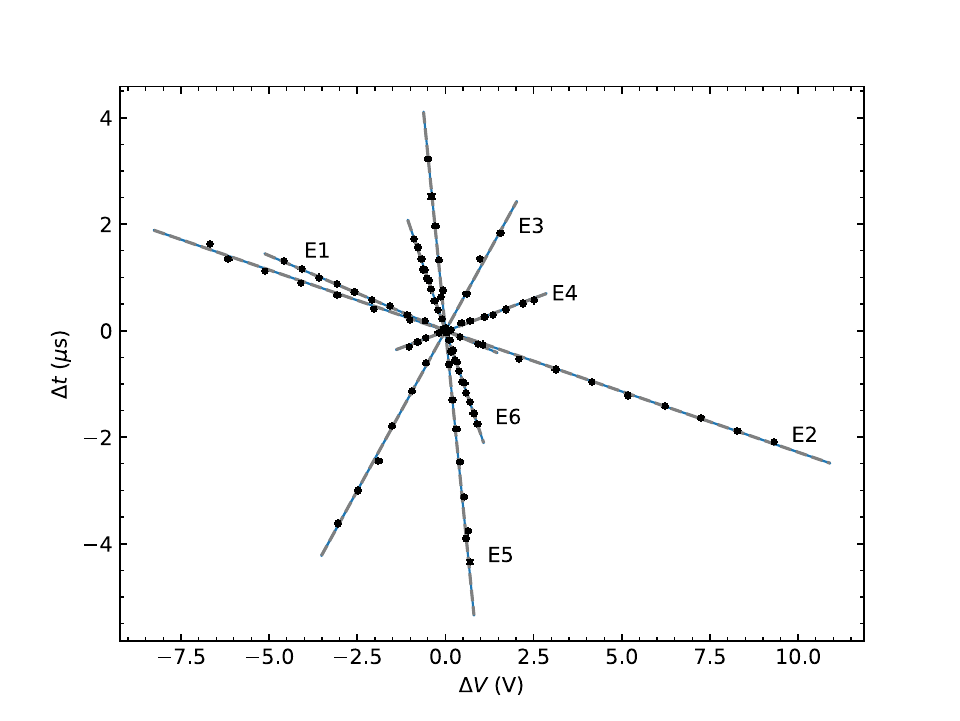}
  \caption{Experimental sensitivity of $^{94}\mathrm{Mo}^+$ ToF at 604 revolutions to changes in the voltages of the MR-ToF-MS. Linear least squares fits with a linear model are drawn on data.}
  \label{fig:mirror_scan}
\end{figure}

\begin{figure*}[th!]
  \centering
  \includegraphics[width=1.0\textwidth]{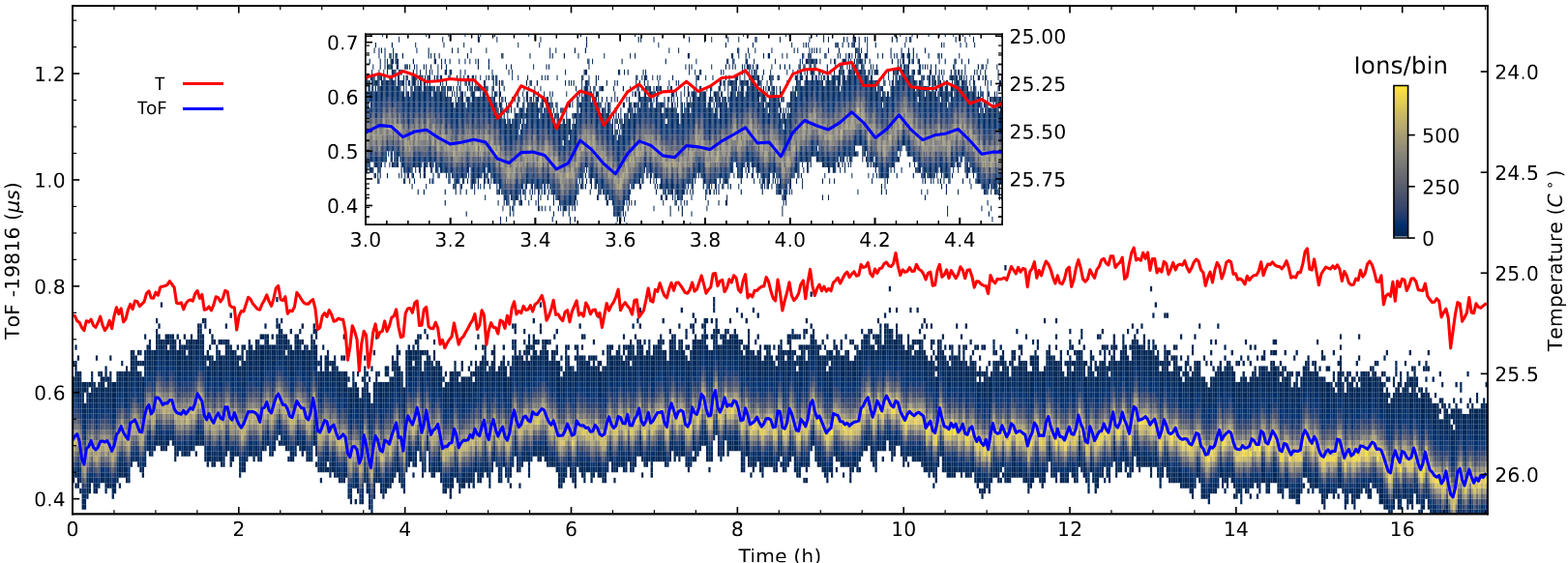}
  \caption{The time-of-flight of $^\mathrm{94}$Mo$^+$ (blue) and temperature of the MR-ToF-MS electrode voltage supply  air cooling exhaust vary (red). The inset displays an enlarged view of a shorter portion of the data, from 3 to 4.5 hours.}
  \label{fig:temp}
\end{figure*}

\begin{figure}[t!]
    \centering
    \includegraphics[width=0.45\textwidth]{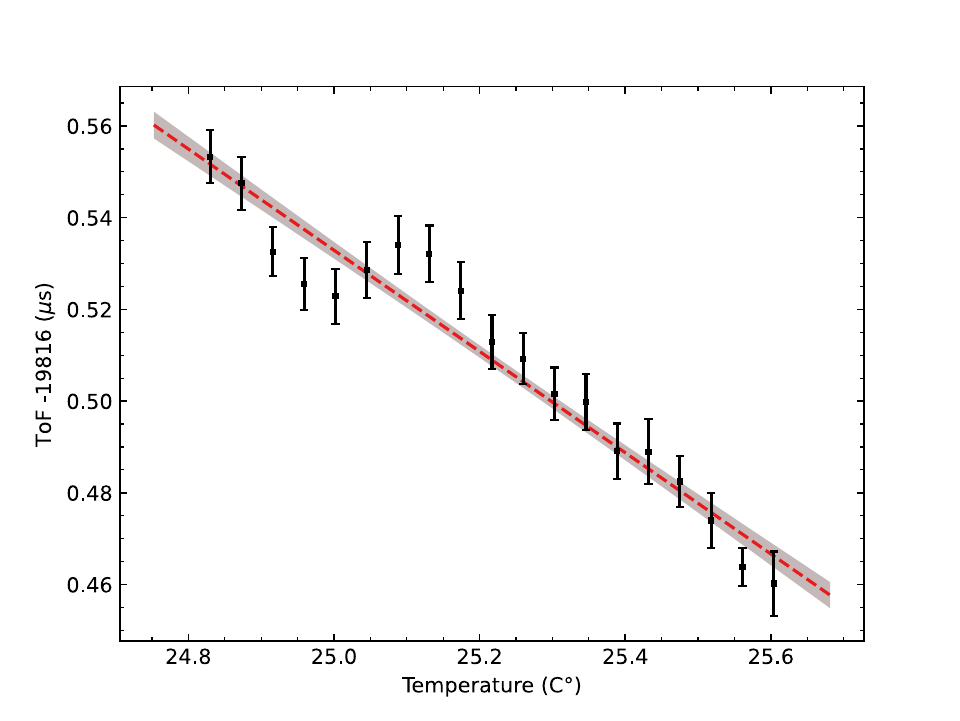}
    \caption{Mean ToF of $^{94}$Mo$^+$ as a function of temperature of the ISEG-EHS 8240X voltage supply exhaust air. A temperature coefficient of $-5.50(30)$~ppm/K was determined from the data.}
    \label{fig:temptofcorr}
\end{figure}

\subsection{Timing sequence and data acquisition}
Ion time-of-flights are constructed from the time differences of the signal of ion impact on a DM291 MagneToF-detector and a "start" timing signal matched with the buncher extraction time. The timings (see Fig.~\ref{fig:timings}) are controlled by two devices: a Spincore PulseBlaster PB24-100-24k-PCI timing card (hereafter PulseBlaster), which controls timings in the trap-line, and a Quantum Composer (QC) that controls the RFQ cooler-buncher timings. The RFQ cooler-buncher, MR-ToF-MS, MagneToF and the TDC are on a high-voltage platform. Therefore, the timing signals from the QC and PulseBlaster are brought from the ground potential to the HV platform over optical fibers using Highland Technologies J720 and J730 converters.  First, the PulseBlaster triggers the QC and commences the RFQ cooler-buncher cycle (see Ref.~\cite{MiniBuncher} for cycle details). Second, the PulseBlaster sends a timing signal to a FAST COMTEC MCS6A time-to-digital converter (TDC) at the ion extraction time from the buncher. While the QC controls the extraction from the RFQ cooler-buncher, the PulseBlaster timing is matched with the extraction. The constant $b$ of Eq.~\ref{eq:tof-eq} includes any remaining offsets. Third, the PulseBlaster signals the pulsed drift-tube (PDT) switch as described in Sect.~\ref{sec:MRToFoper}. Ions extracted out of the MR-ToF-MS hit the MagneToF detector causing the detector to send a signal to the TDC. 

The MCS6A TDC data can be monitored live with an in-house developed user interface that allows to produce histograms of variables such as ToF, detector pulse width, and the time difference of consecutive events. The data are saved event-by-event in binary format and with 200-ps resolution. Both the rising and falling edges of the start and stop signals are saved independently of each other.  External variables, such as laser wavelength and voltages can be saved and correlated with the time-of-flights.

\begin{figure*}[h!]
  \centering
  \includegraphics[width=1.0\textwidth]{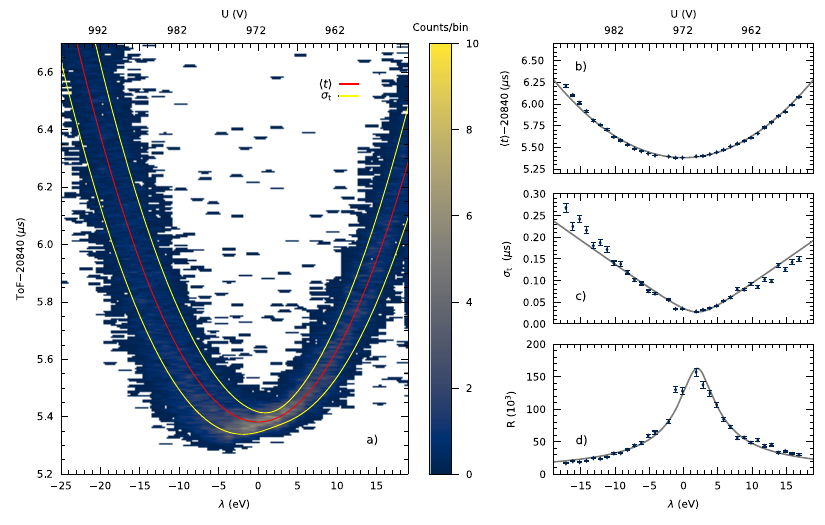}
  \caption{Time-of-flight and derived quantities as a function of pulsed drift-tube voltage. (a) The time-of-flight $t$ of $^{39}$K at 1000 revolutions as a function of parameter $\lambda$. The red and gray lines of (a) and (b) display the fit of $\langle t\rangle(\lambda)$ according to a fit of Eq. (\ref{eq:mean}). The yellow lines show the ToF standard deviation $\sigma_t$. The fit of $\sigma_t$ according to Eq. (\ref{eq:var}) is shown in panel (c). The measured mass-resolving power $R(\lambda)$ is shown in (d) along with fit to Eq. (\ref{eq:mrp}). The highest mass-resolving power is found by trapping the ions with a slightly anisochronous $\lambda \sim 2$V setting.}
  \label{fig:itl}
\end{figure*}


\section{Performance of the IGISOL MR-ToF-MS}
\label{sec:performance}

The IGISOL MR-ToF-MS has been extensively studied with stable ions from the IGISOL off-line ion station \cite{Vilen2020}. In the following sections we summarize its performance and briefly discuss its commissioning in on-line experiments.  
\begin{table}[h!]
    \centering
    \begin{tabular}{lll}
    \hline\hline
    Electrode & Sensitivity ($\mathrm{ppm}/$V) & Uncertainty ($\mathrm{ppm}/$V)\\\hline\hline
    E1 & -14.230 & 0.025 \\
    E2 & -11.513 & 0.009\\
    E3 & 60.60 & 0.09\\
    E4 & 12.48 & 0.10\\
    E5 & -333.62 & 0.29\\
    E6 & -98.12 & 0.09\\ \hline
    \end{tabular}
    \caption{Time-of-flight sensitivity to changes in mirror voltages.}
\label{tab:mirrorV_sens}

\end{table}

\subsection{Sensitivity of ToF to voltages and temperature}
\label{sec:voltage-temporal}

Part per million changes in the voltages of the six electrodes of the MR-ToF-MS affect the ion time-of-flight. The effect of individual voltages on the ToF is displayed in Fig.~\ref{fig:mirror_scan} for $^{94}$Mo$^+$ ions after 604 revolutions in the spectrometer. The outermost electrodes $E5$ and $E6$  have the largest effect on the ToF whereas the inner lens electrodes $E1$ and $E2$ influence the ToF less (see Table~\ref{tab:mirrorV_sens}). This is expected as $E1$ and $E2$ mainly spatially focus the beam while the outermost electrodes are where the ions turn around. The ToF dependence on change in the voltages is very linear; cross-correlations of the voltages' effects on the ToF, however, were not investigated in this study.

The temperature stability of the electrode voltage supply is reported to be 10 ppm/K by the manufacturer. Therefore, a $-4.2$~ppm/K time-of-flight sensitivity is anticipated based on Table \ref{tab:mirrorV_sens} for typical voltages of the IGISOL MR-ToF-MS. The time-of-flight temperature sensitivity was studied with $^{94}$Mo$^+$ ions. Figure~\ref{fig:temp} visualises the ion ToF and the temperature of the ISEG voltage supply exhaust air. The ToF and temperature follow similar trends both on long and short time-scales, which implies a connection between the variables. Figure~\ref{fig:temptofcorr} shows how the measured average ToF relates to temperature. A $-5.55(30)$~ppm/K ToF temperature-sensitivity coefficient was determined from the data. The value is close to the anticipated $-4.2$~ppm/K. Although the temperature of the IGISOL hall does not change drastically over time as seen in Fig.~\ref{fig:temp} - over the course of the 17~h long measurement, the time-of-flight variations are contrained within a 200~ns band - the drifts nevertheless limit the long-term mass-resolving power to $R\lesssim 9\times10^4$. Thus, the data has to be drift-corrected \cite{Wolf2013,Schury2017,Ayet2019} to restore some of the lost resolving power. In the future, stabilizing the temperature or voltage would help to reduce these effects.

\subsection{Energy dependence of ion time-of-flight}

\label{sec:ToF-U-R}
The effect of ion energy on time-of-flight is illustrated in Fig.~ \ref{fig:itl} where the $n=1000$ revolutions time-of-flight of $^{39}$K$^+$ is plotted as a function of the in-trap lift voltage $U$. The red and yellow lines of Fig.~\ref{fig:itl}.a indicate fits to measured $\langle t\rangle$ and $\sigma_t$ from Eqs. (\ref{eq:mean}) and (\ref{eq:var}). The measured $\langle t\rangle$ and $\sigma_t$ are shown in Fig.~\ref{fig:itl}.b and \ref{fig:itl}.c. The locally isochronous point at the ToF minimum was chosen as the reference energy $E_0$. The average time-of-flight $\langle t\rangle$ closely follows Eq.~(\ref{eq:tof_total}). The ToF dispersion $\sigma_t$ minimizes with $\lambda=2$~eV, $\partial t/\partial \lambda>0$ setting confirming that the bunch had de-focused before injection. The fit to Eq. (\ref{eq:var}) describes $\sigma_t$ data fairly well close to $\lambda = 0$ but departs from the data at the edges of the fit as the effect of unaccounted higher-order terms becomes more prominent.  Figure \ref{fig:itl}.d displays the experimental mass-resolving power $R$ fitted with Eq. (\ref{eq:mrp_fancy}). The determined mass-resolving power $R$ closely follows Eq.~\ref{eq:mrp}. The maximum $R\approx 1.5\times10^5$ is reached at approximately $\lambda= 2$~eV.

Mass-resolving power limit from the second-order expansion term $T_2$ and $\sigma_E$ can be estimated by from Eq. (\ref{eq:mrp_fancy}) assuming time-focus (i.e. $(n[2T_2\lambda+T_1]+T_{s1})^2\sigma_E^2 = 0$) at the limit of large $n$, and noting that $T_0$ is much larger than the other terms of the nominator of Eq. (\ref{eq:mrp_fancy}):
\begin{equation}
\label{eq:aber}
    R \approx \frac{T_0}{ 8\sqrt{\ln2}|T_2|\sigma_E^2}.
\end{equation}
The fit indicates $T_2=2.5$~ps/eV$^2$. With $T_0 \approx 20.82$~$\mu s$ and energy spread between 1~eV to 5~eV, Eq. (\ref{eq:aber}) places a mass-resolving power limit to around $5\times10^{4}$--$1.25\times10^6$. Higher limiting $R$ could be reached by reducing  the energy-spread and the magnitude of the second-order expansion coefficient $T_2$. Moreover, a positive $T_2$ produces a pronounced tail on the high time-of-flight side of the bunch, especially when the ions are injected with a mean energy which corresponds to the ToF minimum and any change from the bunch mean energy will result in a higher ToF.

\subsection{Mass-resolving power as a function of revolution number}
\label{sec:tof-n}

\begin{figure}[ht!]
  \centering
  \includegraphics[width=0.43\textwidth]{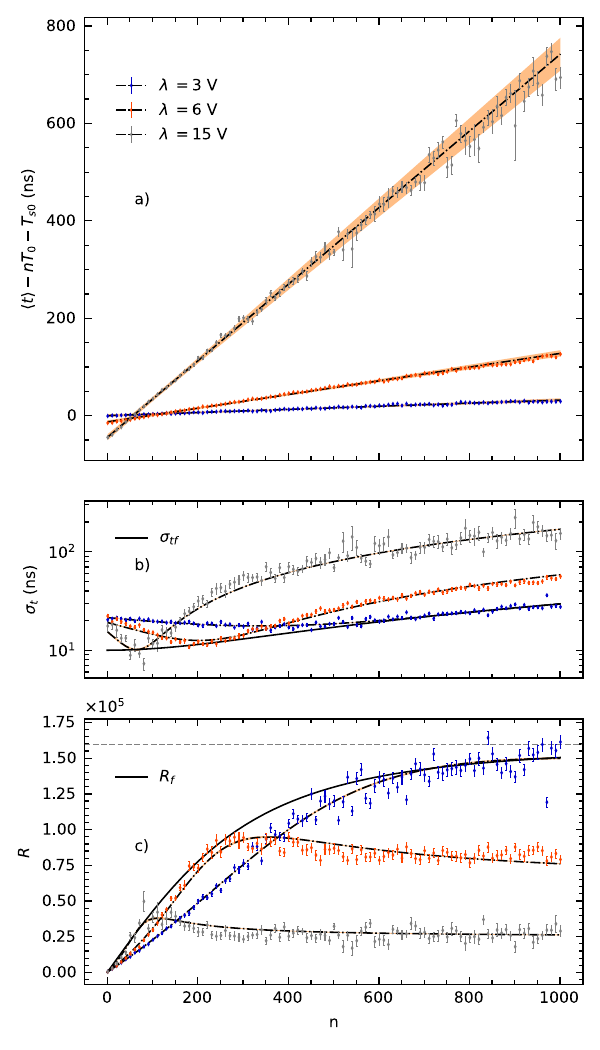}
  \caption{Mass-resolving power of $^{39}$K$^+$ as a function of revolution number $n$ with $\lambda \approx$ 3, 6 and 15~ eV (blue, orange and gray). (a) The non-isochronous part of $\langle t \rangle(n)$. (b) Standard deviation of the temporal distribution $\sigma_t(n)$ (c). The mass-resolving power $R(n)$. The black dash-dotted lines display fits to Eqs. (\ref{eq:mean})-(\ref{eq:mrp_fancy}). The solid black lines are based on Eqs. (\ref{eq:focusedDt}) and (\ref{eq:focused_R}). The dashed line in sub-figure (c) displays the limit of Eq. (\ref{eq:aber}).}
  \label{fig:revs}
\end{figure}

\begin{figure}[h!]
  \centering
  \includegraphics[width=0.45\textwidth]{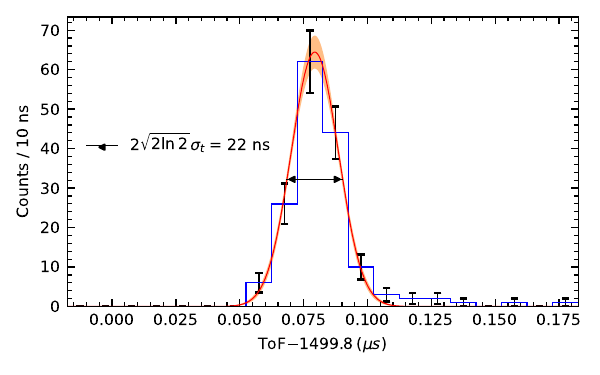}
  \caption{Bunch of $^{39}$K$^+$ in time-focus at 71 revolutions. The red line indicates a Gaussian fit to the data whereas an orange band estimates the model uncertainty. The bunch spans 22~ns in FWHM.}
  \label{fig:bunchwidth}
\end{figure}

The evolution of mass-resolving power as a function of revolution number was studied with $^{39}$K$^+$.  Figure~\ref{fig:revs} shows the mean anisochronous ToF $\langle t\rangle(n)-nT_0-T_{s0}$, the ToF standard deviation $\sigma_t(n)$ and the mass-resolving power $R(n)$ for three settings $\lambda =3,6,$ and 15~eV. The data have been fitted with Eqs. (\ref{eq:mean})-(\ref{eq:mrp_fancy}). The time-of-flight in Fig~\ref{fig:revs}.a increases with revolutions as expected; the ToF increases linearly and at a different rate depending on the trapping energy. Figure \ref{fig:revs}.b indicates the focusing $\partial t/\partial \lambda$ changes depending on the choice of energy as expected from Fig.~\ref{fig:itl}.a. The standard deviation $\sigma_t$ starts from about 20~ns, and decreases as the spectrometer slows down higher-energy ions relative to lower-energy ions. The temporal widths minimize at approximately 70, 200, and 340~revolutions with $\approx$ 10, 13 and 18~ns $\sigma_t$, respectively. A ToF histogram of the first, narrowest focus is shown in Fig.~\ref{fig:bunchwidth} at $n=71$ revolutions. The turnaround time dominates the peak width as the number of revolutions is relatively small. Aberrations limit the width at focus for higher number of revolutions. This can be seen from Eq. (\ref{eq:focusedDt}), where the $\sigma_{t\mathrm{f}}$ at time-focus reads
\begin{equation}
\label{eq:focusedDt}
\sigma_{t\mathrm{f}} = \sqrt{\sigma_\mathrm{th}^2+2n^2T_2^2\sigma_E^2}.
\end{equation}
This curve is drawn in Fig.~\ref{fig:revs}.b. with parameters $T_2 = 2.5$~ps/eV$^2$, from the fit of Fig.~\ref{fig:itl}, and $\sigma_t = 10$~ns and $\sigma_E = 2.8$~eV, which were estimated from the revolution number data. From these values, a longitudinal emittance $\epsilon_\mathrm{long}$ of 175~eVns was estimated for the bunches. The value fits well with the expected 186(10)~eVns emittance, found by scaling the experimentally determined 202(11)~eVns  Mini Buncher longitudinal emittance for $^{45}$Sc$^+$ \cite{MiniBuncher} by the ratio of masses of $^{39}$K and $^{45}$Sc according to Eq. (\ref{eq:emittance}).


Figure~\ref{fig:revs}.c shows the mass-resolving power $R$ as a function of number of revolutions $n$. The peak resolving power shifts to a different number of revolutions by changing $\lambda$. $R\approx 3.8\times 10^4$ is obtained at $n\approx110$ for $\lambda=15$~eV, $R\approx 9.5\times 10^4$ at $n\gtrsim350$ ($\lambda=6$~eV), and $R\approx 1.5\times10^5$ at $n\approx1000$ ($\lambda=3$~eV). Based on Eq. (\ref{eq:focusedDt}), an approximate limiting mass-resolving power $R$ 
\begin{equation}
    \label{eq:focused_R}
    R_\mathrm{f} \approx \frac{nT_0}{4\sqrt{2\ln2}\sqrt{\sigma_\mathrm{th}^2+2n^2T_2^2\sigma_E^2}}
\end{equation}
is drawn, again with the parameters $T_2 = 2.5$~ps/eV$^2$, $\sigma_t = 10$~ns and $\sigma_E = 2.8$~eV (and for simplicity, omitting terms smaller than $T_0$ from the nominator). In terms of Eq. (\ref{eq:aber}), this corresponds to an aberration limit of $R = 1.6 \times 10^{5}$, shown by the gray dashed line. 




\begin{figure}[bh!]
  \centering
  \includegraphics[width=0.45\textwidth]{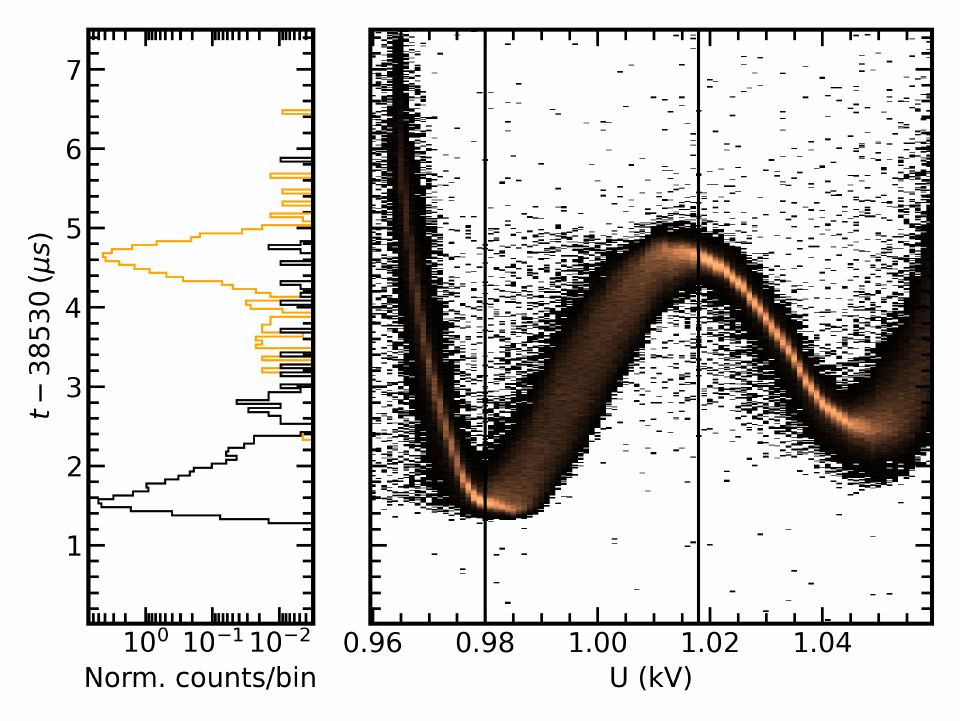}
  \caption{Time-of-flight of $^{133}$Cs as a function of the pulsed drift tube voltage at 1000 revolutions. Projected ToF-slices display a distinct tailing of the ToF distribution, either to higher or lower ToF, depending on the concavity or convexity of $t(E)$.}
  \label{fig:tails}
\end{figure}

A wider scan of the pulsed-drift voltage for 1000 revolutions of $^{133}$ Cs$^{+}$ is displayed in Fig.~\ref{fig:tails}. The scan reveals the effect of higher-order energy terms on the ToF evolution. Nevertheless, Eq. (\ref{eq:tof_total}) can be applied near the three local extrema of $t(E)$ despite being truncated at second-order terms. Notably, the side of the ToF peak tail can be switched from later to earlier ToF by moving to a region for which the local second-order coefficient $T_2$ would be negative instead of positive, as illustrated on the left panel of Fig.~\ref{fig:tails}. This can be useful when an isobaric spectrum containing exotic heavy nuclei is contaminated with stable, lighter nuclei.

\begin{figure}[bh!]
  \centering
  \includegraphics[width=0.45\textwidth]{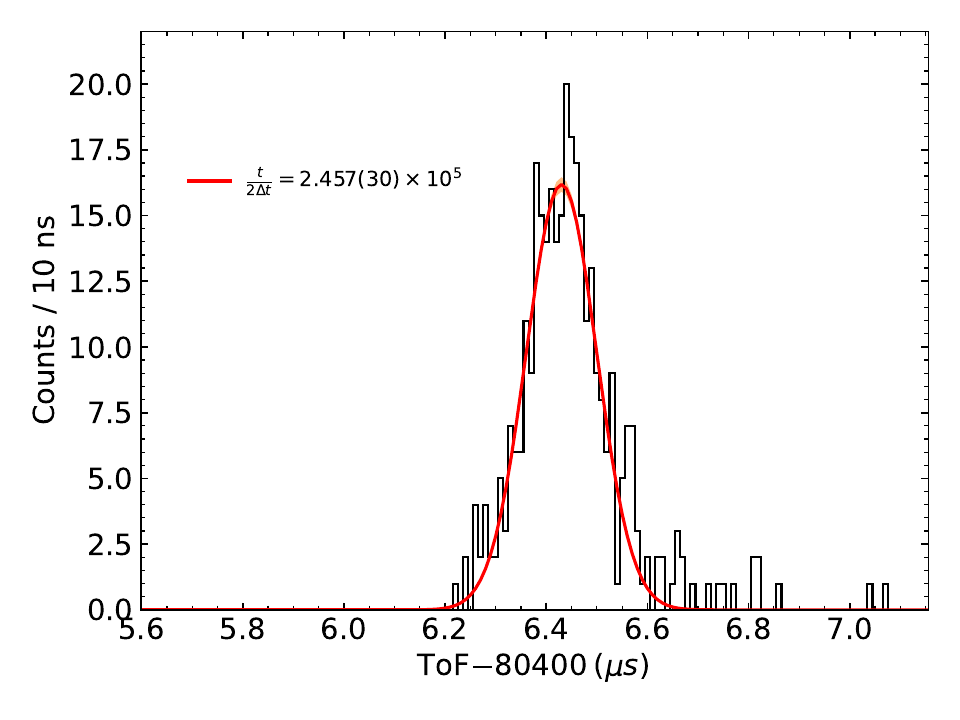}
  \caption{Extraction of the mass-resolving power from the ToF spectrum for $^{133}$Cs$^+$ at 2000 revolutions in the MR-ToF-ms.}
  \label{fig:highmrp}
\end{figure}

Higher mass-resolving powers have been reached with different combinations of mirror voltages and buncher settings. Figure~\ref{fig:highmrp} shows an example of a time-of-flight peak measured for $^{133}$Cs$^+$ ions at $n=2000$ revolutions. The obtained mass resolving-power is $R=2.46(3)\times10^5$; however with this setting the asymmetric tail will always land on the high ToF side. It is likely that the higher resolving power in this particular measurement was caused by an unusually poor transmission efficiency of the beam-line, which resulted in effectively loosing ions from the edges of the energy distribution and narrowing the energy distribution from the usual width. Nevertheless it is clear that the aforementioned 1.6$\times10^{5}$ mass-resolving power is not a hard limit.


\clearpage

\subsection{MR-ToF-MS in on-line experiments at IGISOL}

\begin{figure}[hb!]
  \centering
  \includegraphics[width=0.45\textwidth]{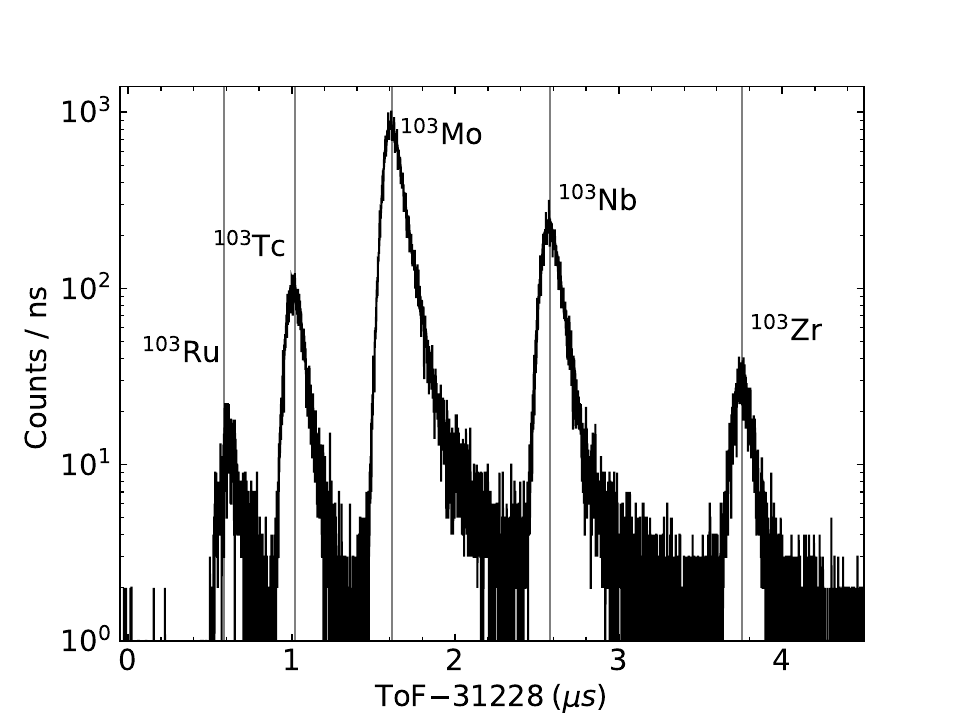}
  \caption{MR-ToF spectrum of the A = 103 isobaric chain. Despite their chemical reactivity, refractory elements such as Mo Nb and Zr can be accessed without significant losses.}
  \label{fig:fission}
\end{figure}

The MR-ToF-MS at IGISOL has been commissioned with radioactive ions in on-line experiments. Measurements of independent fission yields \cite{Korkiamaki2024} and isomeric yield ratios have been performed with it \cite{Cannarozzo2025}. As an example of the quality of the data, Fig.~\ref{fig:fission} shows the time-of-flight spectrum of fission fragments from alpha-induced fission on thorium at mass number $A=103$. The mass-resolving power of the MR-ToF-MS is clearly sufficient to resolve the neighboring ion species. Here the smallest energy difference between the peaks is $Q_{\beta}(^{103}$Tc$)=2.663(10)$~MeV and the largest $Q_{\beta}(^{103}$Zr$)=7.220(10)$~MeV, which require mass-resolving powers of about $R\gtrsim 10^4$ to separate. 

In addition to fission studies, the MR-ToF-MS has been used to detect products in the multinucleon transfer reactions and as a beam composition diagnostic at IGISOL. On-line mass measurements with the IGISOL MR-ToF-MS have been performed with a novel hot cavity laser ion source \cite{Reponen2015}. For example, the masses of the long-lived isomeric states in the $N=Z$ nucleus $^{94}$Ag were measured \cite{Virtanen2025b}. The MR-ToF-MS also served as an ion counter for the laser ionization scheme for $^{94}$Ag, similarly as the JYFLTRAP was used for the laser spectroscopy of $^{99}$Ag in \cite{Reponen2021}. It also has the potential to serve as a fast mass separator for laser and decay spectroscopy experiments, mass spectrometry experiments with the JYFLTRAP Penning trap  \cite{Eronen2012}, and experiments with the MORA setup \cite{Delahaye2019}. 


\section{Summary and Outlook}
The MR-ToF-MS has been successfully commissioned at IGISOL. We have reported on its performance and shown that mass-resolving powers of $\approx 1.5\times 10^5$ can be achieved within 20~ms. The resolving power sensitively depends on the used in-trap lift voltage and reaches its highest value within a few hundred to a thousand revolutions in the MR-ToF-MS. The measurements are in line with the expectation value for the time-of-flight and its variance. 

To enhance the mass-resolving power and the precision of the MR-ToF-MS further, stabilization of the electrode voltages and temperature is needed. Although the effects of time-of-flight drifts can be corrected in software after the measurement, the elimination of this issue at the source should simplify both the measurement and analysis procedure.  Further optimization of the RFQ cooler-buncher and mirror voltages is anticipated to further improve the resolving power. The effect of ion-ion interactions and other systematic uncertainties of the MR-ToF-MS should be carefully studied to mitigate them. To get rid of unwanted contaminant ions, solutions such as re-trapping with a separate Paul trap or a mass range selector in the middle of the drift tube, should be investigated.

The MR-ToF-MS at IGISOL has already been used in many on-line experiments. With the envisaged developments, it will offer a great potential for a range of future experiments at IGISOL. It can be used as a multipurpose instrument at IGISOL, i.e. as a rapid mass spectrometer and beam analyzer, as a fast mass separator for laser spectroscopy, decay spectroscopy and Penning-trap mass measurements, and as an ion counter for yield measurements and laser spectroscopy.

\section*{Acknowledgments}
Funding from the Research Council of Finland under project numbers 273526, 295207, 306980, 327629, 339245 and 354968 and from the European Union’s Horizon 2020 research and innovation program under grant agreement No. 771036 (ERC CoG MAIDEN) is gratefully acknowledged. V.A.V. thanks for the financial support from the Ellen \& Artturi Nyyssönen foundation. Support from the Vilho, Yrjö and Kalle Väisälä foundation is acknowledged by J.R. We thank Prof. Lutz Schweikhard from the University of Greifswald for kindly providing us a simulation package for the MR-ToF-MS.
 \bibliographystyle{elsarticle-num} 
 \bibliography{lahteet,biblio}





\end{document}